\documentstyle[12pt]{article}

\begin{document}
\title{Nonlocal Conservation Laws Derived from an Explicit 
Equivalence Principle}

\author{Rafael A. Vera\thanks{email: rvera@buho.dpi.udec.cl}\\
Deptartamento de Fisica\\
Universidad de Concepcion\\
Casilla 4009. Concepcion. Chile}

\date{October 1997}
\maketitle

\begin{abstract}
According to this principle (EEP), in order that the local physical laws 
cannot change, after changes of velocity and potentials of a measuring 
system, the relativistic changes of every particle and stationary radiation 
(that can be used to measure them) must occur in identical proportion, at 
the same time. In less words, particles and stationary radiations must 
have the same physical properties. Thus in principle better defined 
relativistic laws for particles and their gravitational (G) fields can be 
derived from properties of radiation in stationary state after using 
nonlocal reference frames that don't change in the same way as the 
objects. Effectively, the new laws agree with relativistic quantum 
mechanics and with all of the gravitational tests. The main difference 
with ordinary gravity is linearity. {\it The G field itself has not
energy to exchange with the bodies and it is not a secondary source
of field}.  The EEP also fixes a new cosmic context that has fundamental
differences with the conventional one. This one has been presented in
a separated work as a cosmic test for the EEP. The detailed theory,
that includes the new universe fixed by the EEP, was published in a book.
\end{abstract}

\newpage

\section {The Mayor Source of Errors in Current Literature}

{}From the EEP, the relativistic changes occurring to an object can only 
be detected by nonlocal (NL) observers whose instruments have not 
changed in the same way and proportions as the objects.

This fact is crucial in gravity (G) because the NL experiments such as G 
time dilation and G red shift prove, definitively, that {\it the atoms and 
clocks of observers located in different G potentials are not strictly the 
same with respect to each other, respectively}. Then, strictly, {\it the 
unit systems of observers located in different G potentials are physically 
different compared to each other}.

Hence, {\it most of the current comparisons of quantities measured by 
observers in different G potentials are inhomogeneous, inexact, without 
strict physical meanings}\footnote {Such comparisons can be as 
ambiguous as to compare prices in different countries without previous 
transformations to some common money! This seems to be the mayor 
source of ambiguities and errors in the present century. Hence some 
results of this work may look odd if they are erroneously interpreted in 
terms of current (inhomogeneous) concepts.}.  

Thus to relate quantities measured by observers located in different G 
potentials, they must be previously transformed to a common unit 
system based on a standard (or observer) in some fixed (single) and 
well-defined G potential. To make a difference with the local quantities, 
these transformed (relativistic) quantities have been called {\it nonlocal} 
ones. For the same reason the NL position or potential of the reference
standard has been stated by a subscript. The transformation factors can
be easily found either from NL conservation laws or, approximately, from
strictly homogeneous interpretations of the results of basic experiments,
called {\it gravitational tests}~\cite{v1}$-$\cite{v11}. 

\section{Nonlocal Conservation Laws for Radiation}

\subsection{NL frequency conservation}

Assume that a continuous train of light waves (or periodical light 
signals of a fixed frequency) are emitted at the end $B$ of an optical 
fiber whose other end $A$ is in a higher and constant G potential. From 
{\it wave continuity} it is inferred that not a single wave or light 
signal can disappear during its trip $BA$. According to this, the
theoretical number of them that are crossing the ends $A$ and $B$,
in the same unit of time (like that of the observer's clock at $A$),
must be the same. Then the NL frequency of light traveling 
between $B$ and $A$,{\it with respect to the fixed clock at $A$}, 
cannot change. Then it may be concluded that, in general, {\it the NL 
frequency of light, with respect to a fixed clock (fixed unit system), is 
conserved during its trip throughout static conservative fields} ({\it NL 
frequency conservation law for radiation}.).

Since this property must also holds for single photons, then it may also 
be concluded
 
a) That the G red shift detected by the observer at $A$, of the light 
coming from $B$, cannot possibly be due to a real change occurring 
during the photon's trip. This one can only be due to the lower NL 
eigen-frequencies (lower NL energy  levels) of atoms at $B$ compared 
with those of the same kind of atoms at $A$. Notice that {\it such 
differences do exist long before light emission}. This is most evident in 
G time dilation experiments because the light fly time is negligible 
compared with the measured times and, furthermore, such fly time is 
canceled out after using time differences.

b) That wave continuity must be a property of a single quantum. This 
means that the Huygen's principle can be applied to the {\it 
quantum-wavelets} that should also have some sort of wave continuity.

Thus the differences of the time intervals detected between clocks of 
observers located in different G potentials cannot possibly be due to
any wavelength lost during the light trips between them. They can only
be due to real differences of the natural frequencies of atoms and
clocks located in different G potentials.

Since the photon energy depends only on its NL frequency, then, from NL 
frequency conservation it is also concluded that {\it the relativistic
energy of a photon, with respect to a fixed clock, remains
constant during its free trip throughout conservative fields}. In other
words, {\it a free quantum of radiation cannot exchange energy with
static conservative fields}. This may be called {\it the no energy
exchange law between radiation and G fields}.

Of course, this may eventually look odd because observers at rest in
different G potentials normally measure different frequencies for the 
same light beam. However this is due to the fact that their clocks run
at different frequencies compared to each other. Thus the comparisons 
of frequencies referred to different time units have not well 
defined physical meanings.  On the contrary, such differences can be 
measured just because {\it the NL frequency of the original photons
have remained unchanged during their trips from the NL light source up to 
the local observer position}.

\subsection{Nonlocal Quantum Vector Conservation}

A NL frequency vector oriented in its propagation direction, whose 
absolute value is its NL frequency, can describe more completely the
main NL properties of a photon with respect to the observer. Its
multiple $h$ is called {\it NL quantum vector} because its absolute
value is just the quantum NL energy with respect to such observer.

{}From above, according to wavelet continuity, the NL quantum vector of 
a free photon traveling in a space of isotropic refraction properties 
cannot change. This fact may be called {\it NL  quantum vector 
conservation}\footnote {So far nobody knows the ultimate reasons for
the photon non-spread (stability). Thus we can either accept it, as
a well proved fact, or try to find an explanation. It is reasonable
that the NL refraction index of the space is larger in the regions
of coherent wavelet interference. In this way the photon's wavelets
would be systematically deviated towards the original photon
orientation.}.

\section{Nonlocal Conservation Laws for Particle Models in G Fields}

The simplest (one-dimensional) particle model can be made up of a 
single quantum in stationary state between two perfect mirrors. So far
it is not strictly necessary to know the exact mechanism for
the perfect model reflections\footnote {It is reasonable that
the same mechanism proposed above for the photon non-spread can
account for the critical reflections within particle models.}.

The two components of the model at rest would be mirror reflections 
with respect to each other. Each of them has a half of the quantum 
energy. This may eventually be like a neutrino-antineutrino set. Any 
material part, like an external wave cavity, can be omitted because, 
according to the EEP, their proportional changes must be identical to 
those of the stationary radiation\footnote {Three dimensional models 
with radiation traveling in closed paths can in principle be described by 
the sum of several one-dimensional particle models with different 
orientations and phases between them.}.

Two quantum vectors in opposite directions can represent to the model 
at rest with respect to the observer, each one with a half of the model 
energy.

According to wave continuity, the waves of the quantum wave-trains
(quantum cycles) confined in it are conserved after perfect internal
reflections. Thus, in an isolated model (located in a space of
constant average NL refraction index) the sum of the NL quantum
vectors of and the sum of their absolute values cannot change.  

The same holds for more complex systems made up of may particle 
models. Thus in general, {\it both the sum of the NL quantum vectors
of an isolated system and the sum of their absolute values, with respect
to a fixed clock, cannot change}.

The first sum is the net NL quantum vector of the system. Such vector 
represents to the net number of quantum waves (cycles) per unit of time 
traveling in some well defined orientation of the space. This is called 
the NL quantum vector conservation for isolated systems.

The second sum, that of the absolute values of the quantum vectors, is
called the NL mass-energy of the system.  Such relation is called the
NL mass-energy conservation for isolated systems.

It is simple to verify that the net NL quantum vector of the model is 
equal to the product of its net NL momentum and the NL  speed of 
light~\cite{v3}${-}$\cite{v11}. Thus the conservation of the NL 
quantum vector of a system, in a space of constant average NL refraction 
index, corresponds with the current {\it momentum conservation of 
isolated systems}.

It is obvious that these NL conservation law do correspond 
with the conventional ones but only within local ranges in which the 
current ambiguities can be neglected.

\subsection{Relativistic Quantum Mechanics}

According to the EEP it is not necessary to make additional postulates
to derive, theoretically, its relativistic quantum mechanical properties.
Such properties, like the De Broglie waves, 
come out, naturally, as a consequence of the plain interference of the 
wavelets of the model components~\cite{v4}${-}$\cite{v11}.

Effectively, the NL laws for the particle model do show a complete
correspondence with both special relativity and quantum mechanics.
They make possible to understand the physical phenomena in terms of
the more dual properties of the radiations.

\subsection{Nonlocal Conservation Laws for Particles in G fields}

During a free orbit (or a free fall) of a particle model in a static
G field, its average NL frequency with respect to a fixed unit system
remains constant because the model is made up of radiation that,
according to the no exchange law, cannot exchange energy with the
field (NL mass-energy conservation.

Notice that the model NL momentum changes can only be due to NL 
refraction, i.e., due some gradient of the NL speed of light.
In principle such phenomenon don't change the quantum energy
(self-consistency test). Thus in one way or another, a G field must
have a gradient of its NL refraction index.

The NL mass-energy conservation during a free orbit turns out to be a 
direct consequence of the nature of the particle model. Because, in order 
that the radiation can be confined in the model, the constructive 
interference of the quantum wavelets must occur within the model 
mirrors. Far away from them, the wavelets must interfere with {\it 
random phases}, i.e., destructively. Thus both the net wavelet 
amplitudes and the relative probabilities for existence of energy far away 
from the model must be null. Thus, to the contrary of current beliefs, {\it 
static G fields cannot give up energy to test bodies just because they 
don't have energy}.

Due to its high importance, this basic law has been verified from several 
different ways.  For example, such law can also be derived from the 
current gravitational tests, after using more strictly homogeneous NL 
relationships~\cite{v11}.

Then it is obvious that the energy that appears during the G work can 
only come from the energy confined in test body. This means that
{\it the G work is done at the cost of a decrease of the NL rest
mass-energy of the test body}.  Effectively, after a free fall, it is
evident that such energy is given away just during the stop. Thus the
NL eigen-frequency of either the model or an atom at rest in its final
G potential is smaller with respect to the initial ones. This accounts
for both the G red shift and the G time dilation observed from nonlocal 
viewpoints\footnote {Such changes cannot be detected, locally, at the 
final position, because every local atom has changed in identical 
proportion.}.

\section{The Nonlocal Gravitational field}

The model long range field can only be produced by gradients of the NL
perturbation rate of the space resulting from random phase wavelets 
diverging from it.  This one accounts for the gradients of the NL 
properties of the space, mainly of the NL speed of light and of the NL 
eigen values of the model stationary 
waves~\cite{v4}${\!,\,}$\cite{v5}${\!,\,}$\cite{v11}.

The NL field equation fixed by the EEP is obviously linear. Thus the 
orbits of radiations and particle models turn out to be fixed by plain 
interference of wavelets traveling in a space of variable
NL properties. Thus the model accelerations turn out to come from
gradients of the NL refraction index of the space that in turn induce
gradients of the NL properties of the model. They fix the {\it NL
angular momentum conservation law for radiation and for particles
in G fields}. Effectively, they account for the radar time delays,
G refractions (deviations of light by G fields), planet's orbits and
their perihelion
shifts~\cite{v4}${\!,\,}$\cite{v5}${\!,\,}$\cite{v11}.

{\it The linearity of the G field equation}, which is fixed by the
EEP, results in non conventional properties both for the black holes
and for the universe. For this reason the consistency the new
astrophysical context with the observed facts has been presented as
a test for the EEP~\cite{v11}${-}$\cite{v12}. 

According to the EEP, the experiments for detection of G waves, after 
using round trips of light, must give negative results. Thus such 
experiments would also be fair tests for the present theory provided that 
they can discriminate them, effectively, from any other kind of radiation.

\section{Conclusions}

One of the main advantages of using the EEP as a single base for this 
theory is that the EP is just one of the most unquestionable principles in 
physics. Since it is stated in a more explicit form, it leaves no 
alternatives for making arbitrary assumptions.  This way it takes the 
advantages of the actual knowledge of the dual properties of radiations, 
mainly their quantized properties. In this way everything turns out to be 
ultimately fixed by the most elemental properties of the quantums of
radiation. 

Vice versa, the EEP also makes possible to test the current assumptions 
normally made in current literature and to explore on the nature of 
particles and elemental properties of the radiations.

This work proves that it is possible to get more out from the Equivalence 
Principle, after making it more explicit. Anyway, this work can also be 
used as a guide for future works in this line or for interpreting the 
phenomena observed in the universe. Of course, there is a lot of work to 
do in this direction. This work is just a first and rather small step.

Most of the detailed deductions of this work have been included in a 
book~\cite{v11} that was presented in the Eight Marcel Grossmann 
Meeting (June 1997) so to provide more detailed deductions and 
verifications for the two contributions presented in it.  Such book 
integrates the presentations and publications on this line, done in the 
last 24 years~\cite{v1}${\!-\,}$\cite{v11}. It includes the new 
astrophysical and cosmological context fixed by the EEP.

\end{document}